\documentclass[12pt]{iopart}

\usepackage{graphicx}
\usepackage{amssymb}

\begin{document}

\title{Sympathetic cooling and detection of a hot trapped ion by a cold one}

\author{M Guggemos$^{1,2}$, D Heinrich$^{1,2}$, O A Herrera-Sancho$^{1,3}$, R Blatt$^{1,2}$, C F Roos$^{1,2}$}

\address{$^1$ Institut f\"ur Quantenoptik und Quanteninformation,\"Osterreichische Akademie der Wissenschaften, Technikerstr. 21a, 6020 Innsbruck, Austria}
\address{$^2$ Institut f\"ur Experimentalphysik, Universit\"at Innsbruck, Technikerstr. 25, 6020 Innsbruck, Austria}
\address{$^3$ Centro de Investigaci\'on en Ciencia e Ingenier\'ia de Materiales, Universidad de Costa Rica, 2060 San Pedro, San Jos\'e, Costa Rica}
\ead{christian.roos@uibk.ac.at}
\vspace{10pt}

\begin{abstract}
We investigate the dynamics of an ion sympathetically cooled by another laser-cooled ion or small ion crystal. To this end, we develop simple models of the cooling dynamics in the limit of weak Coulomb interactions. Experimentally, we create a two-ion crystal of Ca$^+$ and Al$^+$ by photo-ionization of neutral atoms produced by laser ablation. We characterize the velocity distribution of the laser-ablated atoms crossing the trap by time-resolved fluorescence spectroscopy. We observe neutral atom velocities much higher than the ones of thermally heated samples and find as a consequence long sympathethic cooling times before crystallization occurs. Our key result is a new technique for detecting the loading of an initially hot ion with energy in the eV range by monitoring the motional state of a Doppler-cooled ion already present in the trap. This technique not only detects the ion but also provides information about the dynamics of the sympathetic cooling process.
\end{abstract}

%
\noindent{\it Keywords}: trapped ions, mixed ion crystals, sympathetic cooling, quantum logic spectroscopy, laser ablation
%
%
%
%

\section{Introduction}
Laser cooling of neutral atoms and ions \cite{Phillips:1998,Leibfried:2003} to temperatures in the 
$\mu K$-range and below has been one key technique enabling the precise quantum control of trapped atomic particles. It has led to spectacular advances in the investigation of quantum degenerate gases, quantum information processing, quantum optics, precision metrology and optical clocks. However, the number of atomic isotopes with a level structure suited for laser cooling is rather small and there are many more atoms and also molecules of interest for experiments at low temperatures that cannot be directly laser-cooled.  

Sympathetic cooling overcomes this limitation by interaction with a second species that can be directly cooled. The Coulomb interaction was the first to be used in two-species non-neutral plasmas of ions \cite{Drullinger:1980,Larson:1986} and in experiments cooling antiprotons by electrons \cite{Gabrielse:1989}. As other interactions can be used as well, the principle of sympathetic cooling is also applied to neutral atoms and molecules \cite{Myatt:1997,deCarvalho:1999}. 

In mixed-species ion crystals, sympathetic cooling has found many applications: it is used for spectroscopy of molecular ions and study of chemical reactions \cite{Drewsen:2004,Hojbjerre:2008,Offenberg:2008,Willitsch:2012} and has been recently used for cooling highly charged ions \cite{Schmoger:2015}. In quantum information processing, ion crystals can be kept cold by cooling one species while preserving quantum information stored in the other species \cite{Barrett:2003,Home:2009}. Sympathetic cooling enables quantum-logic spectroscopy experiments \cite{Schmidt:2005,Hempel:2013} in which one ion species is used to cool and detect another atomic or molecular ion or even an elementary particle \cite{Heinzen:1990}. The low temperatures reachable by sympathetic cooling in these schemes enable the construction of ultra-precise time and frequency standards \cite{Rosenband:2008,Ludlow:2015}.

In small ion crystals, sympathetic cooling has been studied both experimentally \cite{Rohde:2001a,Blinov:2002,Barrett:2003,Lin:2013} and theoretically \cite{Morigi:2001,Hasegawa:2003,Wuebbena:2012}. These investigations were mostly concerned with the cooling limits and the dynamics of the final stage of the cooling process where the ions are already in a crystalline configuration with small oscillations around their equilibrium positions so that the motion can be described by a set of normal modes of motion. By contrast, the initial stage of the cooling process following the loading of an ion has remained largely unexplored. As ion traps provide deep trapping potentials, the energy of an ion loaded into the trap can be in the eV-range and its mean squared distance from the ion which is being directly cooled can be orders of magnitude bigger than its distance in the equilibrium configuration at $T\approx 0$.
As the Coulomb force decreases quadratically with the distance, the energy exchange rate is strongly reduced which can result in long sympathetic cooling times. 

This article reports an investigation of the initial stage of the cooling process. We focus on the most basic situation where a single ion of one species ($^{27}$Al$^+$) is cooled by a single ion of another species ($^{40}$Ca$^+$). We develop a simple model for estimating the time it takes to sympathetically cool down an initially hot ion. We give a description of the experimental apparatus and characterize the velocity distributions of the neutral atoms that are being loaded into the trap by photo-ionization. With a laser-cooled ion loaded into the trap, we show that it is possible to detect the presence of a hot $^{27}$Al$^+$ ion as soon as it is created in the trap and long before the two crystallize. This technique is based on monitoring the vibrational state of the laser-cooled ion by mapping state information to the ion's internal states. The approach not only detects the loading of an ion but provides also information about the sympathetic cooling process.



\section{Modelling the sympathetic cooling dynamics} 
This section presents simple models of sympathetic cooling in two-ion crystals, which provide insights into the cooling dynamics and cooling times. The focus is on the initial cooling stage after loading a hot ion into a Paul trap in which another ion has already been loaded and cooled to low temperatures. 

We assume that the cooling of an ion of mass $m_h$ and charge $e$ by a laser-cooled ion of mass $m_c$ (and charge $e$) can be cast into the following set of differential equations:
\begin{eqnarray}
m_h\left(\ddot{r}_{h,i}+\omega_{h,i}^2r_{h,i}\right)&=&+\frac{e^2}{4\pi\epsilon_0}\frac{r_{h,i}-r_{c,i}}{|\mathbf{r_h}-\mathbf{r_c}|^3}\label{eq:basicmodel_hotion}, \;\; i\in\{x,y,z\}\\
m_c\left(\ddot{r}_{c,i}+\omega_{c,i}^2r_{c,i}\right)&=&-\frac{e^2}{4\pi\epsilon_0}\frac{r_{h,i}-r_{c,i}}{|\mathbf{r_h}-\mathbf{r_c}|^3}+\mathbf{F}_{diss}(\mathbf{r_c},\mathbf{\dot{r}_c}).\label{eq:basicmodel_coldion}
\end{eqnarray}
Here, $\mathbf{r_h}$ ($\mathbf{r_c}$) denotes the position of the initially hot (cold) ion, and $\mathbf{F}_{diss}$ is the friction force of the cooling laser, which, for the sake of simplicity, we assume as being given by $\mathbf{F}_{diss}=-\gamma m_c\mathbf{v_c}$. We model the trapping potential as a static harmonic potential described by oscillation frequencies $\omega_{h,i}$ ($\omega_{c,i}$). While the dynamical confinement by rf-fields in Paul traps can give rise to chaotic two-ion dynamics and heating processes not present in conservative harmonic potentials \cite{Bluemel:1989}, these effects are expected to play a minor role if the initial ion distance is much bigger than the equilibrium ion distance at T=0. 
The Coulomb force appearing at the right-hand side of eqs. (\ref{eq:basicmodel_hotion}) and (\ref{eq:basicmodel_coldion}) couples the motions of the ions. 

At large distances, the interaction between the ions is weak in the sense that for most of the time the Coulomb forces are much smaller than the forces exerted by the trapping potential.  Our approach is therefore to describe the sympathetic cooling process as a sequence of successive Coulomb collisions between the particles which transfer momentum from the hot ion to the one held at low temperature by laser cooling.


For this purpose, it is sufficient to consider small-angle Coulomb collisions: a hot ion passing the cold ion at a minimum distance $b$ with velocity $v$ will experience only a slight deflection of its trajectory. In the absence of a trapping potential, it will exert a force
\[
F\approx\frac{e^2}{4\pi\epsilon_0}\frac{b}{(b^2+(vt)^2)^{3/2}}
\]
on the cold ion and change the momentum of the cold ion by
\begin{equation}
\Delta p = \int_{-\infty}^\infty F(t) dt = \frac{e^2}{4\pi\epsilon_0}\frac{2}{bv} = \frac{e^2}{4\pi\epsilon_0}\frac{2m}{L}, 
\end{equation}
where $m$ is the mass of the hot ion and $L$ its angular momentum. For trapped ions, this approximation is still good if the main momentum transfer happens on a time scale which is short compared to the oscillation period of the ion in the trap. 

An important question is whether the momenta $\Delta p_i$ transferred to the cold ion in consecutive collisions add up constructively or not. In the former case, the energy transferred to the cold ion will be the sum of the momenta squared whereas in the latter case a much bigger energy transfer is possible. It is at this point that the trapping potential plays a crucial role. Because of the trapping force, momentum is not conserved between collisions and the rate at which it changes its sign depends on the mass of the trapped ion. In the following, two models of sympathetic cooling will be presented for the two cases where the ions either have the same or different masses.

\subsection{Sympathetic cooling for ions of different masses \label{sec:coolingmodelsdifferentmass}}
For a pair of ions having different masses, a coherent add-up of momenta is suppressed by the different oscillation frequencies of the ions. We will therefore assume that the energy lost by the hot ion is given by the sum of the energies $E_i=(\Delta p_i)^2 /(2m)$ transferred to the laser-cooled ion. This leads to the following model: we consider a laser-cooled ion of mass $m_c$ and an initially hot one of mass $m_h$ oscillating in the trapping potential with frequency $\omega$ where the trapping potential is assumed to be nearly isotropic with non-degenerate oscillation frequencies. 

As there are two collisions per oscillation period, the rate of energy loss of the hot ion is given by
\begin{equation}
\frac{dE}{dt}=-\frac{\omega}{2\pi}\frac{\langle(\Delta p)^2\rangle}{m_c}
=-\left(\frac{e^2}{4\pi\epsilon_0}\right)^2\frac{m_h^2\omega}{m_c\pi}\left\langle\frac{1}{L^2}\right\rangle
\label{eq:energylossrate}
\end{equation}
Here $\langle .\rangle$ indicates an average over the collision parameters which vary from collision to collision as 
the angular momentum which is not a conserved quantity. To obtain a simple model, we will assume that  $L$ is uniformly  distributed over the allowed range of values, $0\le L\le E/\omega$, and use the rough approximation that $\langle 1/L^2\rangle\approx 4\omega^2/E^2$. After introduction of a length $d$ and an energy scale $E_d$,
\[
  d=\left(\frac{e^2}{2\pi\epsilon_0m_h\omega^2}\right)^{1/3} \mbox{\hspace{.5cm} and \hspace{.5cm}} E_d = \frac{1}{2}m_h\omega^2d^2 = \frac{e^2}{4\pi\epsilon_0 d},
\]
which correspond to the equilibrium distance of a two-ion crystal and the corresponding Coulomb energy of two ions at distance $d$, the differential equation becomes
\begin{equation}
\frac{dE}{dt}=-\left(\frac{\omega}{2\pi}\right)\frac{E_d^3}{E^2}\frac{8m_h}{m_c}
\label{eq:energylossrate2}
\end{equation}
According to this model, the energy decreases as
\begin{equation}
E(t) =\sqrt[3]{E(0)^3-\left(\frac{t}{\tau}\right)\frac{24m_h}{m_c}E_d^3}\label{eq:EnergylossVsTime}
\end{equation}
where $E(0)$ is the initial energy and $\tau=2\pi/\omega$. The time to extract all energy is given by
\begin{equation}
\tau_{cool}=\frac{m_c}{24m_h}\left(\frac{E(0)}{E_d}\right)^3\tau \label{eq:coolingtime}
\end{equation}
with a strong dependence on the initial energy. For example, for an Al$^+$ ion with an initial energy of $E(0)=1$~eV (corresponding to a temperature of $T=E(0)/3k_B \approx 4000$~K in a harmonic trap) oscillating at $\omega=(2\pi)\,1$~MHz, (\ref{eq:coolingtime}) predicts a cooling time of $\tau\sim 5\cdot 10^3$~s whereas for $E(0)=0.1$~eV the cooling time goes down to 5~s. A more refined estimate of $\langle 1/L^2\rangle$ gives rise to a logarithmic correction (see (\ref{eq:refinedcoolingmodel}) in \ref{sec:appendixcoolingmodel}) which slightly reduces the predicted cooling time.

\begin{figure}[tb]
\begin{center}
\includegraphics[width=1\textwidth]{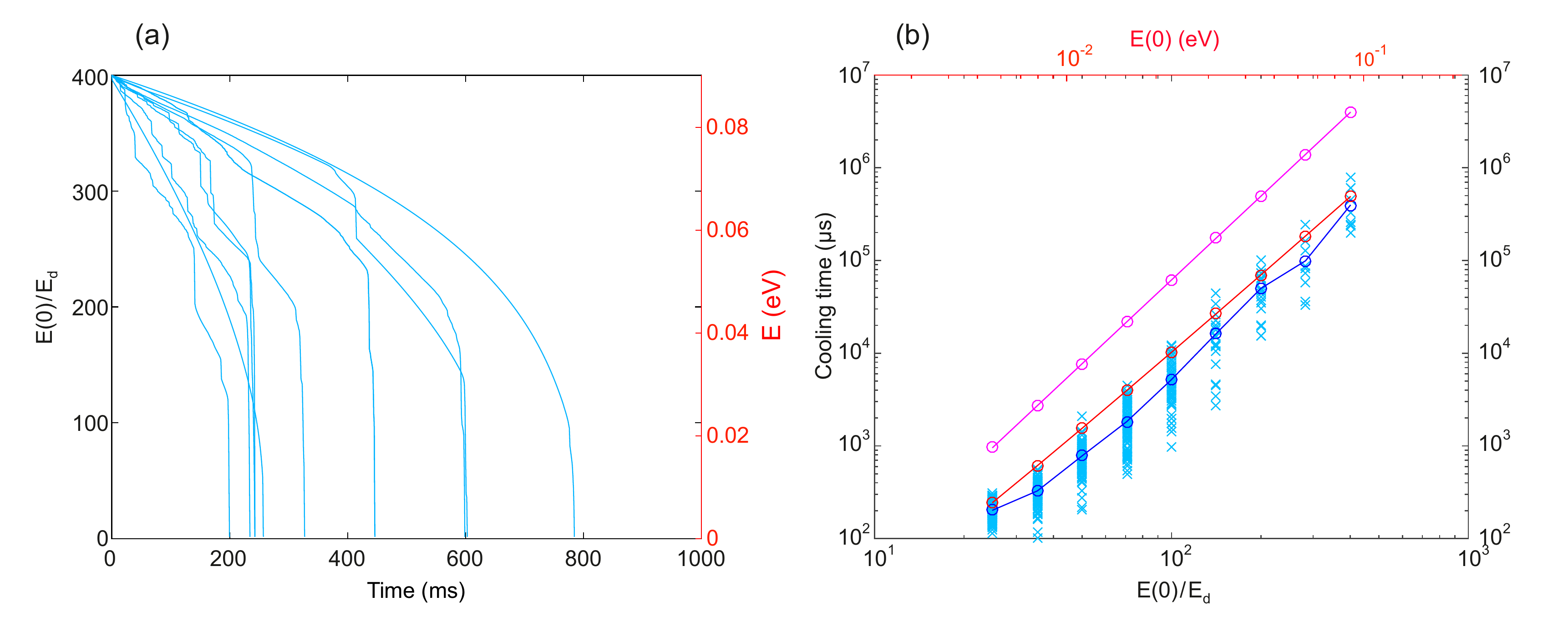}
\caption{\label{fig:SimulationDifferentMassCase}
Numerical simulation of sympathetic cooling of ions of different mass. (a) Numerical integration of the differential equations (\ref{eq:basicmodel_hotion}),(\ref{eq:basicmodel_coldion}). The trajectories show how the total energy of a Ca$^+$-Al$^+$ crystal is lost as a function of time in a nearly isotropic potential with Al$^+$ ion oscillation frequencies of $1.078$~MHz, $1.0563$~MHz, and $1$~MHz along the x-,y-, and z-direction for various initial conditions of the Al$^+$ ion and $\mathbf{r_c}(0)=0$. (b) Cooling times vs initial energy (light blue) from numerical simulations; dark blue circles are the mean cooling time. The pink line is the prediction of (\ref{eq:coolingtime}), the red line the result of the refined cooling model discussed in the appendix (see (\ref{eq:refinedcoolingmodel})). 
}
\end{center}
\end{figure}

Figure \ref{fig:SimulationDifferentMassCase} presents numerical simulations of the cooling process. We use a nearly isotropic trapping potential with slightly incommensurate frequencies which break the rotation symmetry in order to ensure ergodicity. The left panel shows results obtained by numerically integrating eqs.~(\ref{eq:basicmodel_hotion}), (\ref{eq:basicmodel_coldion}) for a potential, in which a hot Al$^+$ ion is placed with a Ca$^+$ ion initially at rest in the centre. The calcium ion experiences a velocity-dependent damping $\sim \gamma v$ with $\gamma = 0.01\omega_z$. Trajectories of the total energy $E(t)$ are shown for different initial conditions of the Al$^+$ ion. The ion's initial position and velocity are randomly drawn by sampling from thermal distributions with temperature $T$ where $E(0)=3k_BT$. For the same initial energy, cooling times may vary by about an order of magnitude. These variations can be partly attributed to different initial energies of the hot ion along the x,y, and z-axis (see \ref{sec:appendixcoolingmodel}). The simulated average cooling time until crystallization occurs is shown in the right panel as a function of the initial energy. A comparison with the predictions of the simple cooling models, see (\ref{eq:coolingtime}) and~(\ref{eq:refinedcoolingmodel}), shows that these models seem to capture most of the physics of the cooling process, in particular the drastic increase of the cooling time with the initial energy. 

It is interesting to consider the dependence of cooling time (\ref{eq:coolingtime}) on the trap frequency $\omega$.  As $E_d\sim\omega^{2/3}$ and $\tau\sim\omega^{-1}$, we have $\tau_{cool}\sim E(0)^3/\omega^3$. If the intial energy $E(0)$ is dominated by the kinetic energy of the neutral atom, the cooling time can be reduced by working with stiff trap confinement. If, on the other hand, the initial energy is dominated by the potential energy that the ion acquires when created off-centre such that $E(0)\sim \omega^2$, then the trap should be made shallow in order to reduce the cooling time.

\subsection{Sympathetic cooling for ions of the same mass\label{sec:coolingmodelssamemass}}
To understand the dynamics of two ions of equal mass in the regime where the Coulomb energy is much smaller than the kinetic and trapping energy, it is useful to split their equations of motion into centre-of-mass and relative motion. Instead of considering the equations of motions of the individual ions, for example along the x-direction,
\begin{eqnarray*}
m\ddot{x}_1&=&-m\omega^2x_1+\frac{e^2(x_1-x_2)}{4\pi\epsilon_0 |r_1-r_2|^3}\\ 
m\ddot{x}_2&=&-m\omega^2x_2-\frac{e^2(x_1-x_2)}{4\pi\epsilon_0 |r_1-r_2|^3}
\end{eqnarray*}
we introduce the centre-of-mass coordinate $\mathbf{r}_c=\frac{\mathbf{r}_1+\mathbf{r}_2}{2}$ and relative coordinate $\mathbf{r}=\mathbf{r}_1-\mathbf{r}_2$ for which we have
\begin{eqnarray}
m\ddot{x}_c&=&-m\omega^2x_c\label{eq:com}\\ 
m\ddot{x}\phantom{_c}&=&-m\omega^2x-\frac{e^2x}{2\pi\epsilon_0 r^3}.\label{eq:stretch}
\end{eqnarray}
In a mean-field approach, the last equation can be approximated by
\begin{equation}
\ddot{x}+\left(\omega^2-\frac{e^2}{2\pi\epsilon_0m}\left\langle\frac{1}{r^3}\right\rangle\right)x=0\label{eq:stretchapprox}
\end{equation}
where $\langle r^{-3}\rangle$ is the average inverse distance cubed between the ions.  Equation~(\ref{eq:stretchapprox}) predicts a reduction of the oscillation frequency to
\begin{equation}
\omega^\prime=\sqrt{\omega^2-\frac{e^2}{2\pi\epsilon_0m}\left\langle\frac{1}{r^3}\right\rangle}
\end{equation}
because of the Coulomb repulsion between the ions and, in conjunction with~(\ref{eq:com}), a coherent exchange of the energy between the ions with an angular frequency
\begin{equation}
\Delta\omega=\omega-\omega^\prime\approx\frac{e^2}{4\pi\epsilon_0m\omega}\left\langle\frac{1}{r^3}\right\rangle=\frac{\omega}{2}\left\langle\frac{d^3}{r^3}\right\rangle.
\end{equation}
Therefore, in the absence of dissipation, the energy exchange time from the hot to the cold ion scales as $(E/E_d)^{3/2}$ and can be much faster than the energy transfer between ions of different mass. Neglecting the position averaging and setting $E(0)=m\omega^2r^2$, the exchange time becomes
\begin{equation}
\tau_{ex}=\frac{\tau}{\sqrt{8}}\left(\frac{E(0)}{E_d}\right)^{3/2}.
\end{equation}
For $E(0)=1$~eV, $m=27$~u, and $\omega=(2\pi)\,1$~MHz, the energy exchange can happen in a time of $\tau_{ex}\approx 0.1$~s, orders of magnitude faster as compared to the case of ions of different mass.

\begin{figure}[tb]
\begin{center}
\includegraphics[width=1\textwidth]{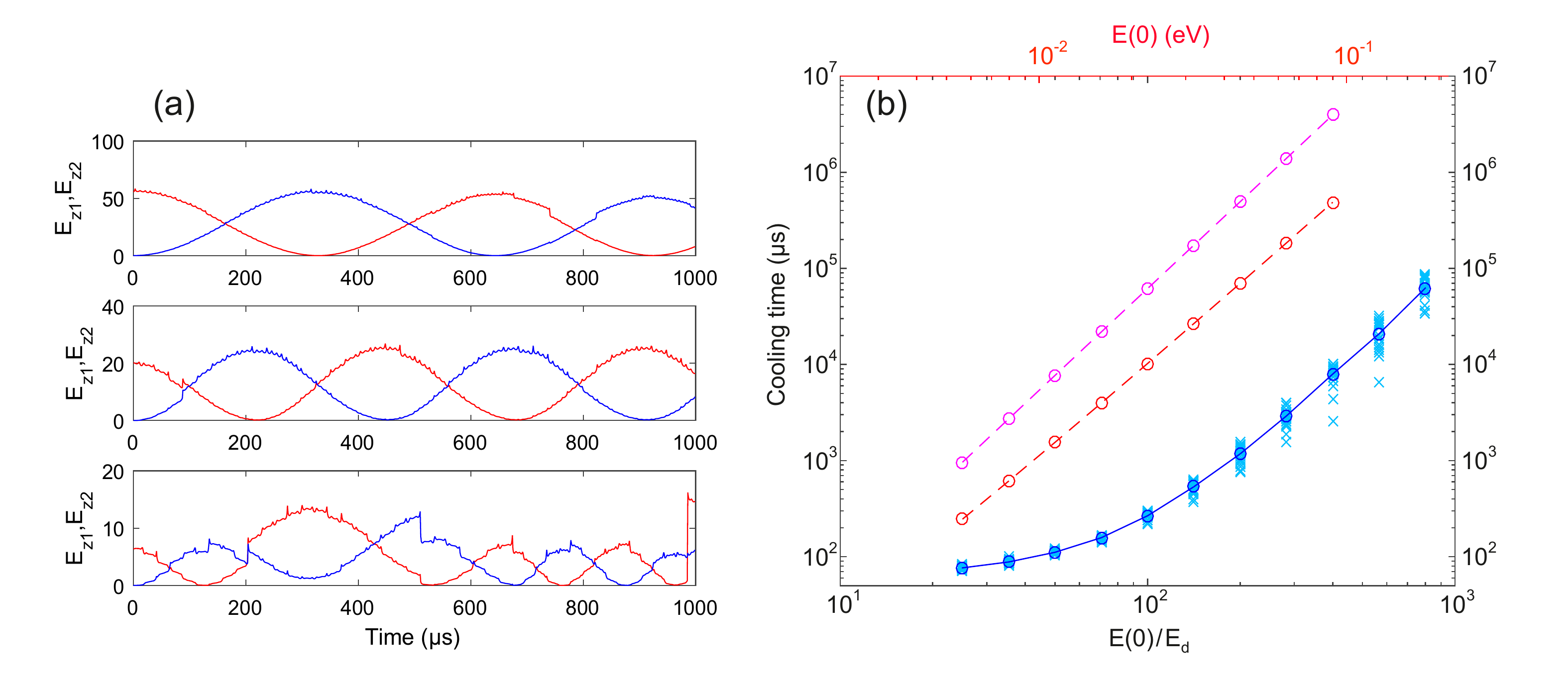}
\caption{\label{fig:SimulationSameMassCase}
Numerical simulation of sympathetic cooling of ions of equal mass. Numerical integration of the differential equations (\ref{eq:basicmodel_hotion}),(\ref{eq:basicmodel_coldion}) for a pair of ions of equal mass. All parameters are the same as in Figure~\ref{fig:SimulationDifferentMassCase} except for the mass of the cooling ion ($m=27u$). (a) In the absence of a friction force, the ions resonantly exchange their energies along the z-direction with a frequency which depends on the total initial energy (here shown for $E(0)/E_d=100$) and the repartition of energy along x,y,z. A similar energy exchange is also observed along the x- and y-direction.  (b) Cooling times vs. initial energy (light blue) from numerical simulations; dark blue circles are the mean cooling time. The pink line is again the prediction of~(\ref{eq:coolingtime}), the red line the result of the refined cooling model discussed in the appendix (see~(\ref{eq:refinedcoolingmodel})). The crystallization time is more than an order of magnitude shorter than for ions of unequal mass. 
}
\end{center}
\end{figure}

Numerical simulations of the full two-ion dynamics confirm that a coherent exchange of energy between the ions indeed happens. Figure~\ref{fig:SimulationSameMassCase}(a) shows the exchange of potential and kinetic energy along the z-direction in the absence of a friction force. For the depicted dynamics, the total energy is $E=100\,E_d$. Sudden jumps in the energy are caused by Coulomb collisions redistributing energy among the different directions $x$, $y$, and $z$. The energy exchange frequency may change depending on the repartition of the energy (see \ref{sec:appendixcoolingmodel}). 
 For friction forces with $\gamma>0$, the dissipation will suppress the coherent energy exchange if $\gamma$ is substantially bigger than the exchange rate. For this reason, the same scaling of the cooling time with initial energy $E(0)$ as for the case of unequal masses is to be expected. However, Figure~\ref{fig:SimulationSameMassCase}(b) shows that nevertheless much shorter cooling times can occur as a result of a partial coherent addition of momentum transfer by subsequent Coulomb collisions.

\section{Experimental apparatus}

The experimental apparatus has been developed for quantum-logic spectroscopy experiments \cite{Schmidt:2005,Wuebbena:2012} using $^{40}$Ca$^+$ as the logic and $^{27}$Al$^+$ as the spectroscopy ion. For the sympathetic cooling experiment presented in this paper, we cool and coherently manipulate Ca$^+$ ions whereas Al$^+$ serves as an ion species of different mass which is sympathetically cooled.

\begin{figure}[tb]
\begin{center}
\includegraphics[width=0.6\textwidth]{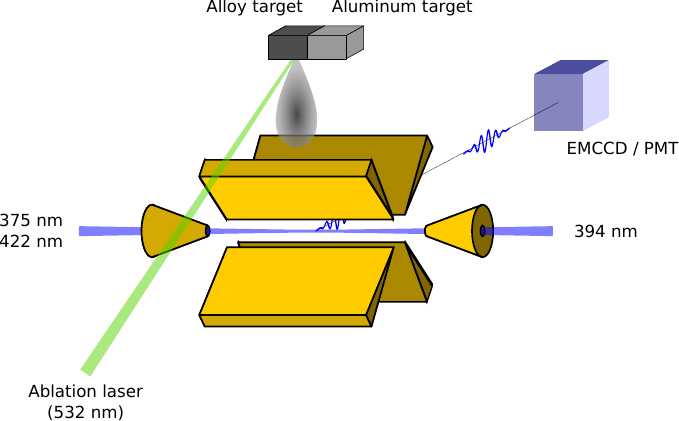}
\caption{\label{fig:setup}
Schematic of the experimental setup: Ions are trapped in a linear Paul trap. Their fluorescence is detected by an EMCCD camera and/or photomultiplier tube. Two ablation targets are arranged side by side above the trap. Single pulses of a laser at 532~nm ablate neutral atoms from one of the targets. Atoms crossing the trap centre may be ionized and trapped. The photo-ionization laser beams are sent along the trap axis and pass through the trap's endcaps which have holes of  0.5~mm diameter. 
In some experiments, the photomultiplier is also used to record blue fluorescence photons emitted by neutral calcium and aluminium atoms crossing the trapping volume.
}
\end{center}
\end{figure}

The experimental setup shown in Figure~\ref{fig:setup} consists of a linear Paul trap in an ultra-high vacuum system. The trap electrodes are made of titanium with a 10~$\mu$m thick layer of gold on top and are mounted in a sapphire holder. The ion to rf-blade distance is about 500~$\mu$m, the distance between the endcaps is 4.5~mm. Holes in the endcaps allow for optical access along the trap axis. For the experiments described below, calcium ions are confined with typical trap frequencies of about 400~kHz in the axial and 2~MHz in the radial direction. A pair of coils creates a magnetic field of 3~G, required for laser cooling and optical pumping,  along the axial direction.

Four different lasers are used for laser cooling and coherent manipulation of the calcium ions.
Light of a frequency-doubled extended-cavity diode laser at 397~nm Doppler-cools the ion on the S$_{1/2}$ to P$_{1/2}$ transition and optically prepares it in one of the Zeeman S$_{1/2}$ states. Further extended-cavity diode lasers emitting at 866~nm and 854~nm pump out the D$_{3/2}$ and D$_{5/2}$ metastable states. For coherent manipulation of the $S_{1/2}\leftrightarrow D_{5/2}$ quadrupole transition, an ultra-stable 
Ti:sapphire laser with a linewidth of ~1Hz is used.

Fluorescence photons emitted by Ca$^+$ ions at 397~nm are collected by an objective, which covers 2.5\% of the full solid angle and is installed outside the vacuum chamber. By means of moveable mirrors and beam splitters, the light collected by the objective can be sent either to a photomultiplier tube, to an EMCCD camera or to both detectors simultaneously. The imaging system covers an area of about $330\cdot 330\,\mu$m$^2$ at the trap centre. In addition to detecting Ca$^+$ fluorescence photons, it can also be used to measure the fluorescence of neutral Ca and Al emitted at wavelengths of 423~nm and 394~nm.

Loading of aluminium ions by photo-ionization requires a beam of neutral Al atoms crossing the trapping volume. As thermal sources of aluminium have to be heated to high temperatures to create a sufficient flux of atoms, we opted for creating neutral atoms by short-pulse laser ablation from a target \cite{Hendricks:2007,Leibrandt:2007,Sheridan:2011}. Two ablation targets are installed next to each other at a distance of 26~mm from the trap in a direction perpendicular to the trap axis. One target consists of pure aluminium, the other one is an alloy composed of 70\%~Al and 30\%~Ca used for creating an atomic beam of calcium.

For laser ablation a frequency-doubled Nd:YAG laser at 532~nm with a pulse duration of 2~ns and a maximum pulse energy of 400~$\mu$J is used. The energy of the pulses can be attenuated by means of a half wave plate and a polarizing beam splitter.  A second half wave plate and a calcite beam displacer allow for individually addressing both targets. 
Both beams are focused onto the targets with a 200~$\mu$m full-width half maximum (FWHM) diameter onto the alloy and 50~$\mu$m FWHM onto the Al target. The beams hit the targets under a grazing angle of 30$^\circ$. For loading the ion trap, pulse energies of about 100~$\mu J$ are used.

Calcium and aluminium ions are loaded into the trap by two-step photo-ionization \cite{Gulde:2001}. Resonant excitation to an excited state of Ca (Al) is achieved by extended-cavity diode lasers emitting at 423~nm (394~nm). For the second excitation step, a free-running diode laser at 375~nm is used for Ca. For Al, the laser at 394~nm has sufficient energy for coupling the excited state to the continuum states.

\section{Experimental results}

\begin{figure}[tb]
\begin{center}
\includegraphics[width=0.7\textwidth]{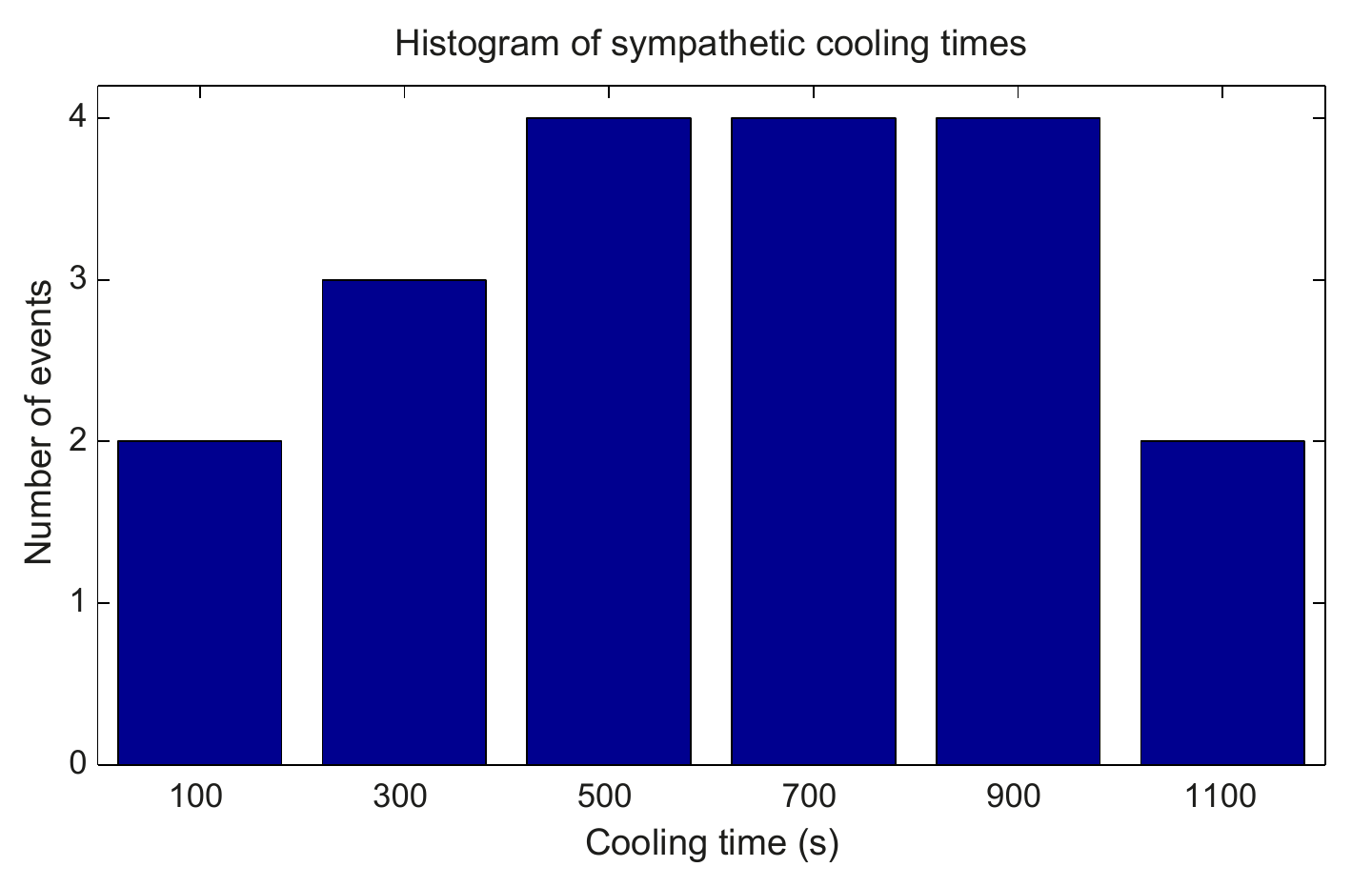}
\caption{\label{fig:AlCoolingTimeDistribution}
Distribution of sympathetic cooling times, measured by recording the time it takes a laser-cooled Ca$^+$ ion to crystallize a hot Al$^+$ ion, starting from the instant of laser ablation. The histogram (with bin size 200~s) shows that the cooling times may vary by an order of magnitude.
}
\end{center}
\end{figure}

Ablation loading of ions by single ablation pulses enables a precise measurement of the duration of time between the instance when an ion is loaded into the trap and the moment when it crystallizes. The measurements discussed in this paper were sparked by the observation that sympathetic cooling of an Al$^+$ ion by a Ca$^+$ ion could take minutes before two ions crystallized. These unexpectedly long cooling times initially proved a impediment to deterministic loading of Ca$^+$/Al$^+$ two-ion crystals because they limited the repetition rate at which Al loading trials could be carried out. Figure~\ref{fig:AlCoolingTimeDistribution} shows the cooling time distribution obtained from 19 successful trials of loading an Al$^+$ ion into a trap housing a cold Ca$^+$ ion. The mean cooling time till crystallization occurs is about $\tau\approx 600$~s.
Each loading trial should have a probability $p$ substantially smaller than one of loading a single Al$^+$ ion in order to avoid loading more than a single ion by an ablation laser pulse. Under these conditions, the loading time of a mixed crystal $\sim \tau/p$ became painfully long. To overcome this  limitation, we developed a method which detects hot Al$^+$ ions by the disturbance they cause on the motional state of the Ca$^+$ ion.

\subsection{Cooling dynamics of Al$^+$ ions sympathetically cooled by Ca$^+$}
A hot Al$^+$ ion oscillating in the trap exerts a quasi-randomly fluctuating Coulomb force on the cold Ca$^+$ ion. From this perspective, the Al$^+$ ion, once it gets loaded into the trap, acts as a high-temperature reservoir heating the motion of the Ca$^+$ ion.  A sensitive method of detecting the ion therefore is to cool the Ca$^+$ ion to a low temperature, let it interact with the hot ion and to measure its motional state using an atomic transition in the resolved-sideband regime. 

This kind of approach has been widely used in quantum information processing experiments for investigation of electric-field noise sources caused by fluctuating voltages on the trap electrodes \cite{Brownnutt:2014}. For this purpose, an ion is ground state cooled and after a waiting time excited on the upper or lower vibrational sideband. A measurement of the excitation probability to an electronically excited state reveals information about the motional state as the electronic coupling strength depends as $\propto\sqrt{n+1}$ or $\propto\sqrt{n}$ on the motional excitation $n$.

To estimate the size of the heating effect, we can express the initial energy loss rate predicted by~(\ref{eq:energylossrate2}) in terms of the number of phonons gained per second. For an initial energy of $E=1$~eV, the formula predicts a heating rate of about $d\langle n\rangle/dt\sim 4E_d^3/(3hE^2)\sim 3700$ phonons/s, which is much higher than the ambient heating rate of our trap and which  would be easily detectable by sideband spectroscopy. 

However, given the expected rate, there is not even a need for carrying out spectroscopy on a motional sideband. Instead, the dependence of the coupling strength $\Omega_{nn}$ on the carrier transition on the motional state of a vibrational mode, $\Omega_{nn}=\Omega_0\langle n|e^{i\eta(a+a^\dagger)}|n\rangle$, provides a means of unambigouosly mapping the average motional quantum number $\bar{n}$ on an electronic excitation probability. As shown in \ref{sec:appendixcouplingstrength}, we expect the average squared coupling strength to be given by
\[
\langle\Omega^2\rangle_{\bar{n}}
\approx \Omega_0^2\exp\left(-2\eta^2\bar{n}\right)I_0\left(2\eta^2\bar{n}\right)
\]
for an ion in a thermal state of motion. Therefore, the following pulsed measurement scheme (repeated many, many times) can be used to detect the aluminium ion:
\begin{enumerate}
\item The Ca$^+$ ion is Doppler-cooled into the Lamb-Dicke regime and prepared in the $S_{1/2}$ ground state by optically pumping out its metastable D states.
\item During a waiting time $\tau$, the ion is heated if an Al$^+$ is present in the trap.
\item A $\pi$-pulse is carried out on the $S_{1/2}\leftrightarrow D_{5/2}$ transition which deterministically transfers the ion into the $D_{5/2}$ state in the absence of an Al$^+$ ion. In the presence of an Al$^+$ ion, the reduced transfer probability $p_{\bar{n}} \approx \sin^2\left(\frac{\pi}{2}\sqrt{\langle \Omega^2\rangle_{\bar{n}}}\right)$ is indicative of the energy of the Al$^+$ ion.
\item A fluorescence measurement detects whether the Ca$^+$ was transferred into the $D_{5/2}$ state.
\end{enumerate}
%

\subsubsection{Experimental implementation}
For an experimental demonstration of the scheme described above, a calcium ion was loaded into a trap with trap frequencies of about  $\nu_{x,y}^{Ca}\approx 2$~MHz and $\nu_z^{Ca}\approx 0.4$~MHz.
In each experimental cycle, the trapped Ca$^+$ ion was Doppler-cooled for 5~ms.
After a waiting time of 10~ms, two $\pi$-pulses were carried out on the 
$|S_{1/2},m=-1/2\rangle\leftrightarrow |D_{5/2},m=-3/2\rangle$  and $|S_{1/2},m=1/2\rangle\leftrightarrow |D_{5/2},m=3/2\rangle$ transitions
  and the state of the Ca$^+$ was detected in a 5~ms long fluorescence measurement on the $S_{1/2}\leftrightarrow P_{1/2}$ transition. This measurement sequence was repeated about 50 times per second in order to estimate the excitation probability to the $D_{5/2}$ state. Then, in one experimental cycle, the photo-ionization beam at 394~nm was switched on and a single ablation pulse with a pulse energy of about 100~$\mu$J was fired on the target with the goal of creating a single Al$^+$ ion in the trap. After the ablation, the series of experimental cycles continued in order to detect whether the loading trial was successful.

\begin{figure}[tb]
\begin{center}
\includegraphics[width=0.9\textwidth]{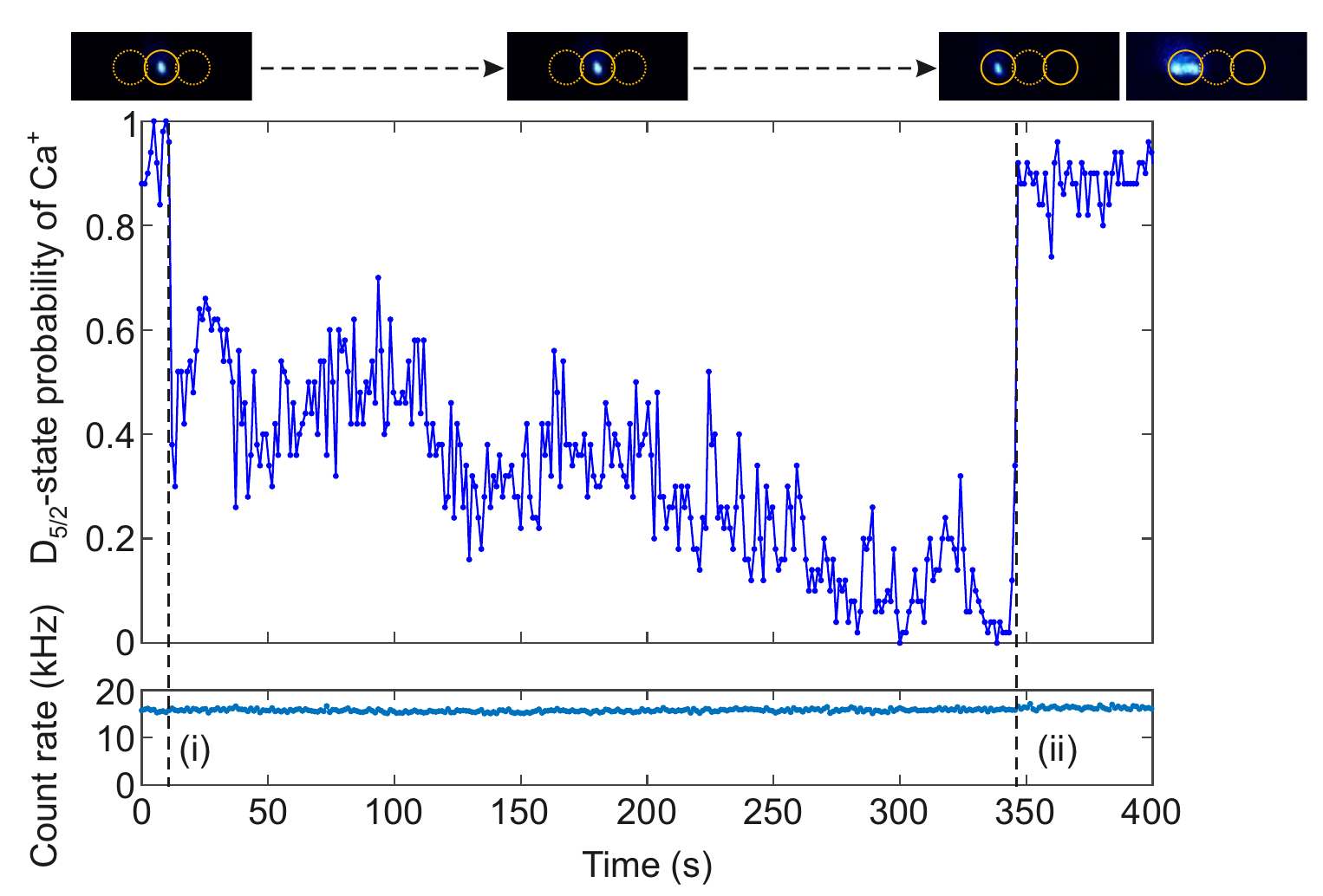}
\caption{\label{fig:AlDetection}
Detecting the loading of a hot ion by a cold one. After 10~s, a hot Al$^+$ ion gets captured (i) in a trap loaded with a laser-cooled Ca$^+$ ion. The Ca$^+$ ion is heated by the hot Al$^+$ ion and as a result, the probability of transferring the calcium ion into its $D_{5/2}$ state by a $\pi$-pulse drops markedly (upper plot). Over the next 340~s, sympathetic cooling via the Ca$^+$ reduces the energy of the Al$^+$ ion until it finally crystallizes (ii). At this point, Doppler cooling efficiently cools the resulting two-ion crystal and the $\pi$-pulse transfer efficiency jumps up again. The loading of the Al$^+$ ion is not detectable in the fluorescence rate of Ca$^+$ ion (lower plot), nor can it be detected by spatially resolving the fluorescence on a CCD camera. The pictures show the position of the Ca$^+$ ion over the course of the measurement. The circles indicate the position of the single-ion Ca$^+$ crystal and the positions of Ca$^+$ and Al$^+$ in the two-ion crystal. The rightmost picture shows the motional excitation of the two-ion crystal by modulating the axial potential at the frequency of the axial in-phase mode for a measurement of the mass of the dark Al$^+$ ion.
}
\end{center}
\end{figure}

The measurements shown in Figure~\ref{fig:AlDetection} demonstrates the power of the technique. In this plot, the calcium ion's quantum state detected in sets of 50 consecutive cycles is used to estimate the excitation probability $p_{\bar{n}}(t)$ as a function of time. The loading of the Al$^+$ ion manifests itself as a marked drop of $p_{\bar{n}}(t)$. During sympathetic cooling, there is a trend to lower and lower excitation probabilities as the cooling progresses. It is caused by the increasing heating rate the calcium ion experiences till the moment when the aluminium ion crystallizes and manifests its presence on the CCD camera by shifting the equilibrium position of the calcium ion away from the centre of the trap.

\begin{figure}[tb]
\begin{center}
\includegraphics[width=0.9\textwidth]{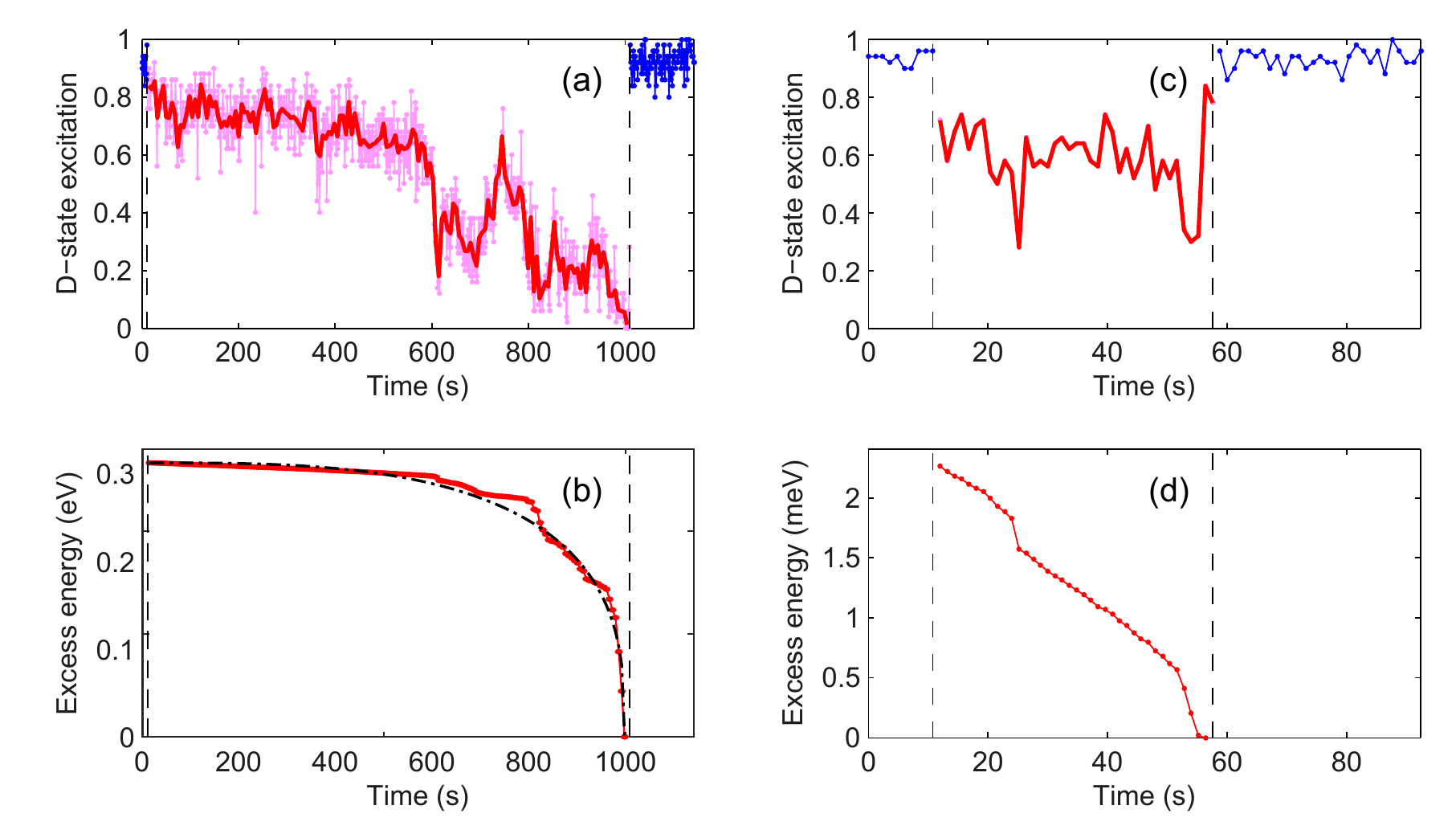}
\caption{\label{fig:SympatheticCoolingRate}
Sympathetic cooling dynamics of an $^{27}$Al$^+$ ion cooled by a $^{40}$Ca$^+$ ion. (a) $D_{5/2}$-state excitation probability after a $\pi$-pulse. The first dashed vertical line marks the moment when an Al$^+$ ion is created by an ablation pulse. The second line indicates the moment of crystallization. The estimated $D_{5/2}$-state excitation shown as red (pink) dots between the line results from binning 250 (50) experimental repetitions together. (b) Energy of the Al$^+$ ion deduced from the binned data in (a). The dash-dotted line is the temporal energy loss predicted by~(\ref{eq:EnergylossVsTime}) with initial energy and cooling time adjusted to match the experimental signal. (c) Same type of signal as in (a). In this case, the Al$^+$ ion was transiently trapped for about 40~s before getting lost again. (d) The energy extracted by sympathic cooling is two orders of magnitude smaller as compared to the cooling event shown in the left panels. 
}
\end{center}
\end{figure}

\begin{figure}[tb]
\begin{center}
\includegraphics[width=0.7\textwidth]{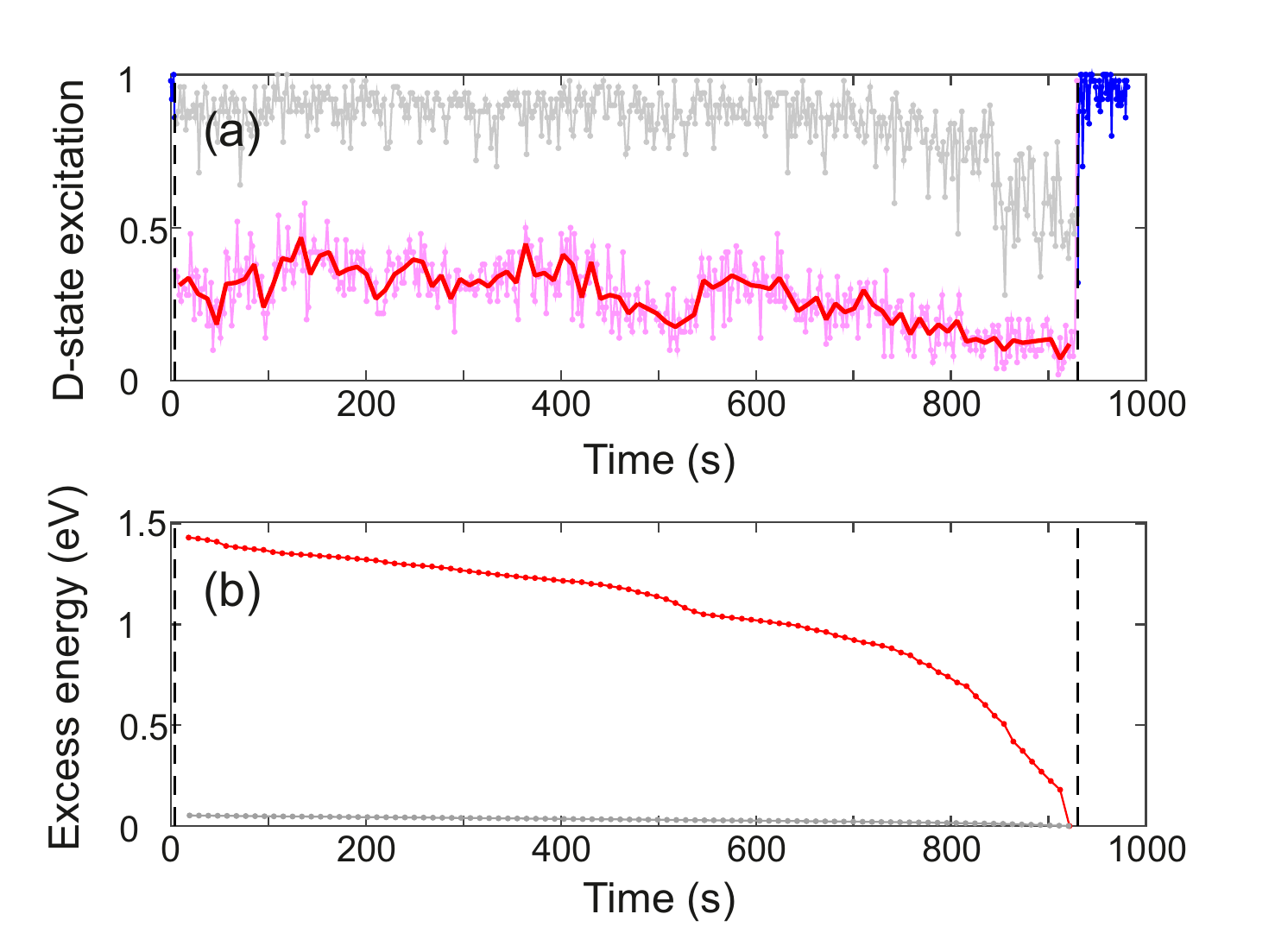}
\caption{\label{fig:SympatheticCoolingRate_secondfig}
Sympathetic cooling dynamics of an $^{27}$Al$^+$ ion cooled by a $^{40}$Ca$^+$ ion with the same type of curves as in Figure~\ref{fig:SympatheticCoolingRate}. In this experiment, the transfer probability to the D-state was measured by omitting the waiting time for every second data point (grey line). These control experiments show that the transfer probability stays high till the final stage of the cooling process. (b) Excess energy inferred from the experiments with waiting time. The error made by assuming that the calcium ion has zero temperature at the beginning of the waiting time is very slight (grey line). 
}
\end{center}
\end{figure}

As the excitation probability is a function of the motional state of the Ca$^+$ ion at the time of the coherent excitation, it provides information about the rate at which the ion is heated by the Al$^+$ ion. The rate is equal to the energy loss rate of the latter. This dependence enables an estimation of the amount of energy the Al$^+$ ion had when it was created in the trap. Figure~\ref{fig:SympatheticCoolingRate}(a) shows another sympathetic cooling event monitored by measuring the excitation probability $p_{\bar{n}}(t)$ of the Ca$^+$ ion to the $D_{5/2}$ state. By converting $p_{\bar{n}}(t)$ into a heating rate which is subsequently integrated over time, the excess energy of the Al$^+$ ion with respect to the energy it had at the time of crystallization can be deduced (see panel (b)). The energy loss as a function of time shows features also predicted by the simple cooling model presented in section \ref{sec:coolingmodelsdifferentmass}: following an intial slow loss of energy, the bulk of the energy is dissipated in the last cooling stage before crystallization occurs. The measured energy of $0.3$~eV rests on the assumption that the heating rate by the Al$^+$ ion is spatially isotropic. A second assumption being made is that laser cooling instantaneously dissipates all the energy transferred to the Ca$^+$ ion so that its vibrational state stays close to the Lamb-Dicke regime even in the final stage of the cooling process. This assumption can be tested by interleaving experiments with and without waiting time before the $\pi$-pulse and comparing the resulting electronic excitations. In such experiments one of which is shown in Figure~\ref{fig:SympatheticCoolingRate_secondfig}, we observe that laser cooling maintains the temperature until the final stage of the sympathetic cooling process where a slight but insignificant rise of the temperature can be observed. For the conversion of $p_{\bar{n}}$ to a heating rate, we binned 250 experiments to reduce the influence of quantum projection noise as this type of noise results in overestimates of the actual heating rate.  

The detection of Al$^+$ ions by monitoring the induced heating rate on a Ca$^+$ ion also reveals that in some cases Al$^+$ ions were transiently loaded without ever being cooled to low temperatures.  Figure~\ref{fig:SympatheticCoolingRate}~(c)--(d) shows such an event in which the Al$^+$ ion was lost from the trap after about a minute. A possible explanation for such losses, which in some cases happened even many minutes after the capture event, could be that following a close encounter with the cooling ion, the energy gets repartitioned among the three spatial directions and the ion escapes along a direction where the potential well is not deep enough to confine it.   

\subsection{Cooling dynamics of Ca$^+$ ions sympathetically cooled by Ca$^+$ ions}
Instead of detecting Al$^+$, the detection method described before can also be applied to infer the prescence of a hot Ca$^+$ ion. In this case, the hot Ca$^+$ ion is cooled both sympathetically and directly by laser cooling. In the regime where the Doppler shift of the ion is much bigger than the detuning of the cooling laser from resonance, simple models of laser cooling predict a cooling rate that scales only weakly  with the energy of the ion, $dE/dt~\sim 1/\sqrt{E}$ \cite{Wesenberg:2007}. It is therefore not obvious that sympathetic cooling could be the dominant cooling process. However, numerical estimates based on the model developed in~\cite{Wesenberg:2007} result in cooling times of many seconds to minutes for $E_0=0.1\ldots 1$~eV if the cooling laser is set to conditions optimal for Doppler cooling. The long time scales are confirmed by our experiments: we are not able to cool an ion created by laser ablation to low temperatures when the parameters of the Doppler-cooling laser are optimized for achieving low temperatures. Therefore  the direct laser cooling of the hot ion is negligible in the sympathetic cooling experiments described in the following.

\begin{figure}[tb]
\begin{center}
\includegraphics[width=0.7\textwidth]{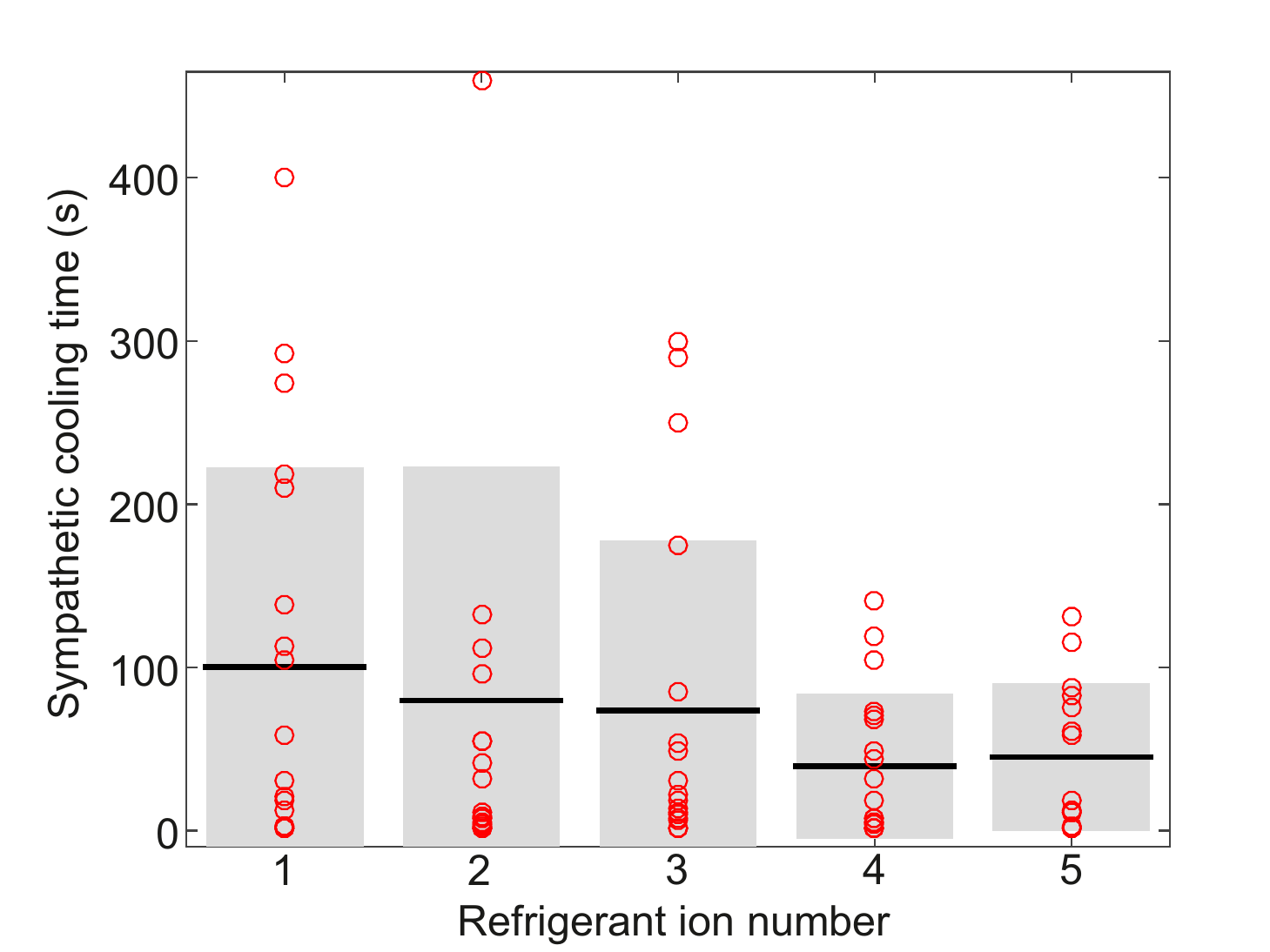}
\caption{\label{fig:CaCoolingTimeDistribution}
Distribution of sympathetic cooling times of a Ca$^+$ ion as a function of the number $n$ of refrigerant ions. The red circles show the individual cooling times for each experiment. The mean values for each data series are indicated by the black lines with calculated standard deviation error bars (grey rectangles). Assuming that the initial energy of the hot ion is identical on average for each data series, a $\sim1/n$ dependence of the sympathetic cooling time on the refrigerant ion number is expected. The notably shorter cooling times of column 2 may be explained by the small sample size.
}
\end{center}
\end{figure}

Figure~\ref{fig:CaCoolingTimeDistribution} summarizes the sympathetic cooling times observed when cooling Ca$^+$ ions. We find  a mean cooling time of about 3 minutes, shorter than the times observed for Al$^+$ but considerably longer than what we would expect from the numerical simulations of the cooling process for ions of equal mass. Interestingly, the observed cooling times vary by more than two orders of magnitude. One hypothesis that could explain the long cooling times is that ions with a high initial energy might explore regions of the trap with a sufficiently high anharmonicity such that the resonant energy exchange postulated in section \ref{sec:coolingmodelssamemass} for ions of the same mass becomes suppressed as the oscillation frequency of the hot ion is no longer equal to the one of the cold ion. In our setup, anharmonicities are strongest along the axial direction of the trap. Simulations of the trapping potential show that resonant exchange of axial energy between two Ca$^+$ ions will be suppressed for oscillation amplitudes bigger than 80~$\mu$m.

We also carried out sympathetic cooling measurements with more than one refrigerant ion. In this case, one would expect a reduction of the cooling time scaling inversely with the number of ions acting as a cold reservoir. As shown in Figure~\ref{fig:CaCoolingTimeDistribution}, we indeed observe a marked reduction of the cooling. Due to the limited number of experiments, the data is too noisy to make precise statements about the scaling behaviour. However, the observed cooling time shortening is yet another indication that the dominant cooling mechanics is sympathetic cooling and not direct laser cooling. 

In the experiments presented in this manuscript, sympathetic cooling is more efficient than direct laser cooling of hot ions. However, this statement should not be interpreted to be valid in general for arbitrary detunings of the cooling laser. For ions with high kinetic energy, scaling arguments show that direct laser cooling (with the laser detuning optimized for the initial energy) should win over sympathetic cooling.

For both Al$^+$ and Ca$^+$ ions, the measured sympathetic cooling times are much longer than anticipated, pointing to the ions having fairly high initial energies when captured in the trap. One contribution to this energy is the velocity of the neutral atoms that are photo-ionized. To investigate whether the kinetic energies of laser-ablated atoms are substantially higher than the one emanating from a thermal oven, we characterized the velocity distributions of the laser-ablated neutral atoms as described in the next section.

\subsection{Velocity distributions of laser-ablated Ca and Al atoms}
We characterized the velocities of ablated Al and Ca atoms by time-resolved fluorescence spectroscopy of the neutral atoms crossing the trapping volume. For these measurements, we triggered single ablation laser pulses and measured the fluorescence of laser-ablated atoms induced by the laser at a wavelength of 394~nm (423~nm), which excites the respective species to the intermediate state in the photo-ionization process. The fluorescence was recorded with sub-10~ns resolution by the photomultiplier tube that normally detects the fluorescence of trapped calcium ions at 397~nm. 
The photons detected by the photomultiplier were converted into TTL pulses and recorded on a fast oscilloscope triggered by the light of the ablation laser pulse which was recorded by a fast photodiode. The photon detection time $\tau$ is inversely proportional to the atomic velocity, $v=d/\tau$, where $d=26$~mm is the distance travelled by the atoms from the ablation target to the detection region in the centre of the trap.

\begin{figure}[tb]
\begin{center}
\includegraphics[width=0.9\textwidth]{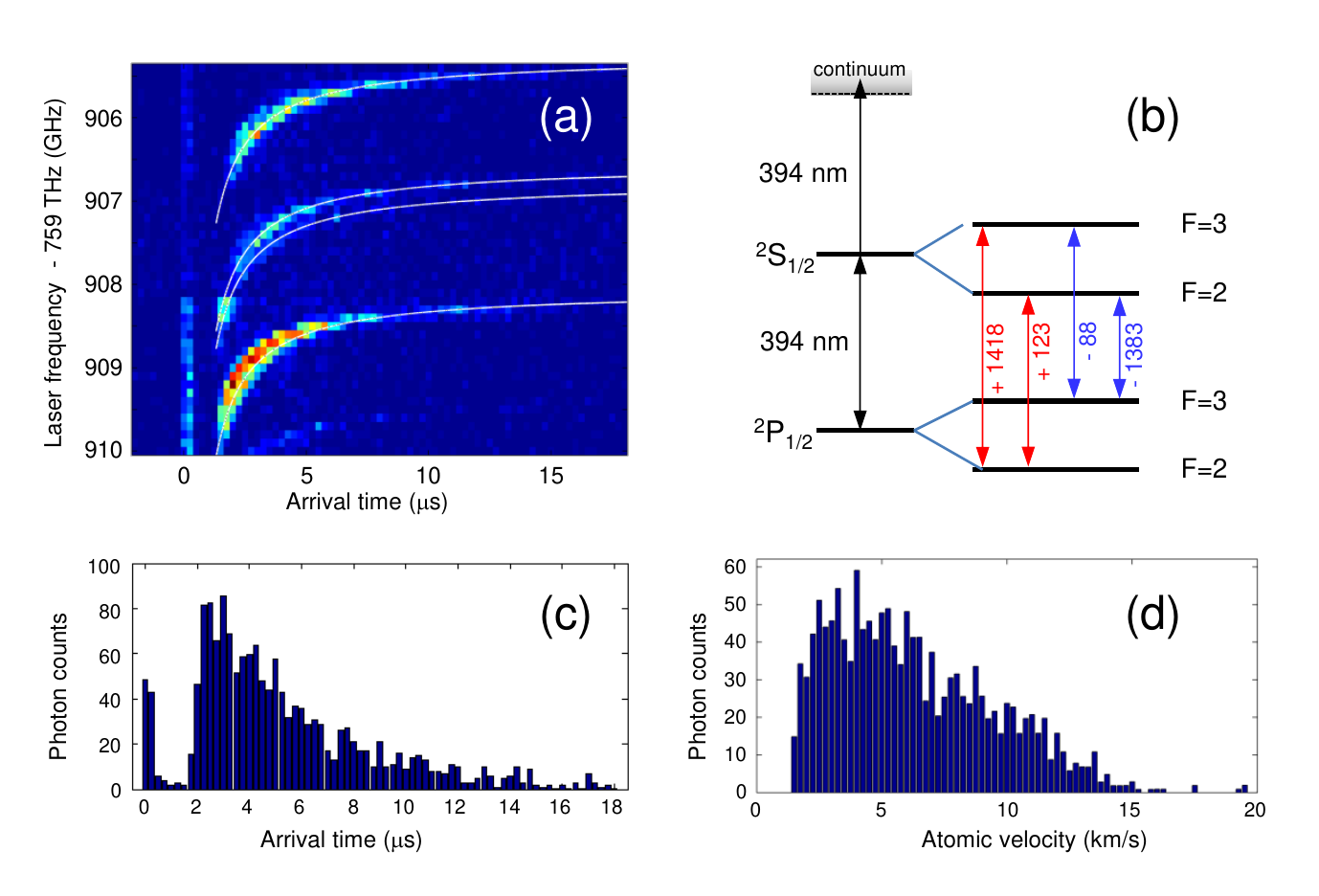}
\caption{\label{AlSpectroscopy}
Time-resolved fluorescence spectroscopy of the $^2P_{3/2}\leftrightarrow {{^2}S}_{1/2}$ transition.  (a) Time-resolved fluorescence of Al atoms crossing the centre of the trap versus laser detuning. The number of detected photons is colour-coded in the image. The vertical line at $\tau\approx 0$ results from photons produced by the laser ablation pulse. The white lines indicate the expected resonances for an angle between the atom and the laser of 2.3$^\circ$. (b) Hyperfine structure of the transition with hyperfine shifts of the transition frequency \cite{Nakai:2007} in units of MHz. (c) Histogram of atomic arrival times. (d) Histogram of atomic velocities. For further information, see main text.  
}
\end{center}
\end{figure}

For each value of the photo-ionization laser frequency, the ablation laser was triggered five times with pulse energies of $200~ \mu$J. Figure \ref{AlSpectroscopy}(a) shows the time-resolved fluorescence of aluminium atoms on the $^2P_{3/2}\leftrightarrow{^{2}S_{1/2}}$ transition at 394~nm. In order to cover the four hyperfine transitions, the laser was scanned over a range of more than 4~GHz in steps of 200~MHz. The hyperfine resonances show a velocity-dependent frequency shift that is consistent with the assumption that the atomic beam is not perfectly perpendicular to the photo-ionization laser but spans an angle of  $87.7^\circ$.

To get rid of Doppler shifts, the detected photon counts stemming from one of the hyperfine transitions can be summed up in frequency space. Figure~\ref{AlSpectroscopy}(c) shows the frequency-summed arrival time distribution measured on the $F=3\leftrightarrow F^\prime=2$ transition after subtraction of background counts. The data shows an initial burst of photons, which is independent of the frequency of the photo-ionization laser. The signal occurs in the first 500~ns after the ablation pulse. The physical mechanism causing the signal is currently unclear. One hypothesis is that electrons created by the ablation pulse might hit the trap electrodes and create the fluorescence photons. About $2\,\mu$s after the ablation pulse, the fastest atoms arrive with velocities in excess of 10~km/s. We can transform the arrival time distribution into an atomic velocity distribution by assuming that every atom scatters $\sim 1$ photon before being pumped into either another hyperfine state of the $^2$P$_{1/2}$ or the $^2$P$_{3/2}$ state independent of its velocity. To obtain the velocity distribution, we convert the recorded times $\tau_i$ of photocounts into velocities via $v_i=d/\tau_i$ and build a histogram by binning the $v_i$ into velocity classes. From this histogram, we subtract the effect of the PMT background count rate based on a measurement of the background count rate measured by counting photons arriving at times $\tau<0$ prior to the ablation pulse. The resulting distribution displayed in Figure~\ref{AlSpectroscopy}(d) shows that the distribution peaks at velocities close to 4.5~km/s. These measurements confirm much earlier observations \cite{Dreyfus:1986,Gilgenbach:1991} that the velocity distribution cannot be explained by a Maxwell-Boltzmann distributed flux of atoms with a temperature equal to that of the heated laser-ablation target. We checked that the velocity distribution measured at $200\,\mu J$ pulse energy was not significantly different from the one used for loading ions.

We would expect that only ions created from the low-velocity tail of the distribution can be trapped in our experiments. This assumption can be tested in experiments in which the photo-ionization laser intensity is switched on only after a delay time $\tau_d$ with respect to the ablation laser pulse or pulsed on in a time interval $\tau_d+[0,\,\tau]$. In the experiments, we did not succeed in loading atoms for $\tau_d\gtrsim32\,\mu$s, corresponding to atomic velocities of $v\lesssim 800$~m/s. Likewise, we found that we were not able to capture and trap atoms when $\tau_d+\tau\lesssim 7\,\mu$s, corresponding to atomic velocities of $v\gtrsim$~3700~m/s. While the lower velocity limit is given by the low density of slow atoms, the upper velocity is likely to given by the capture velocity of the trap, which is set by the depth of the trap.

\begin{figure}[tb]
\begin{center}
\includegraphics[width=0.9\textwidth]{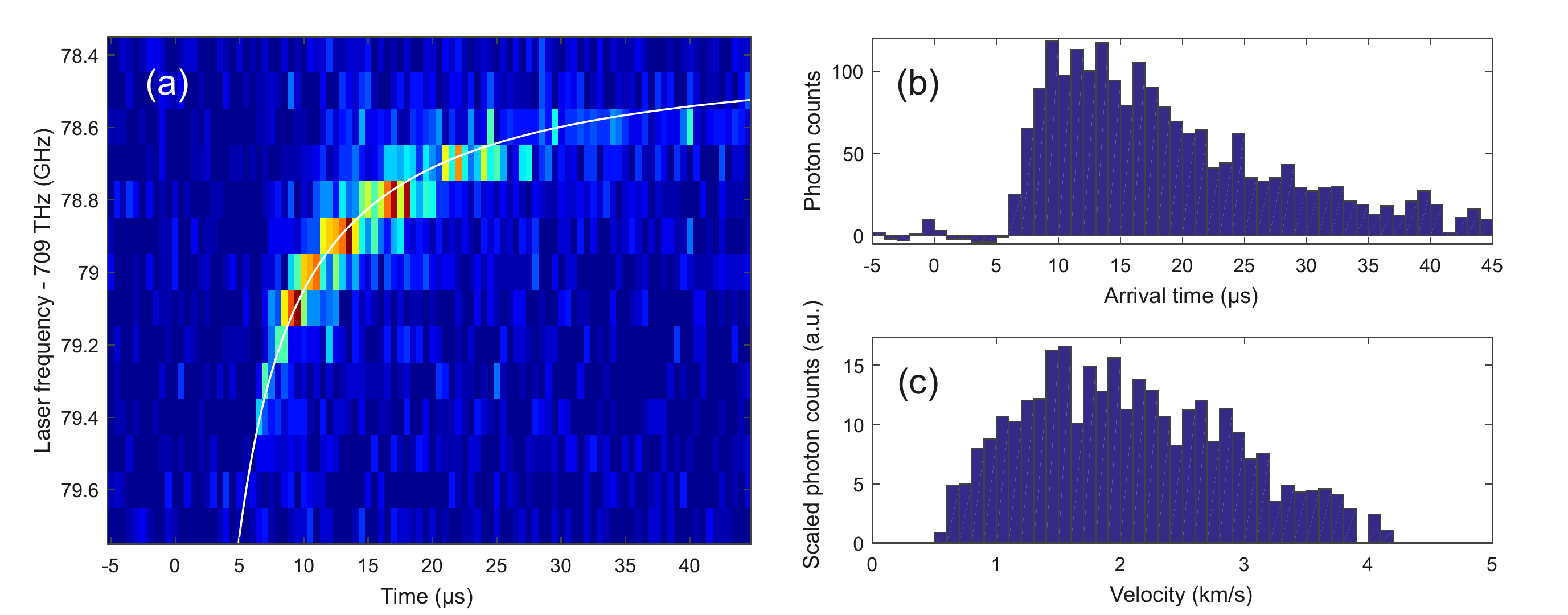}
\caption{\label{CaSpectroscopy}
(a) Time-resolved fluorescence spectroscopy of the $^1S_0\leftrightarrow {{^1}P}_1$ transition of  $^{40}$Ca atoms crossing the centre of the trap. The white line indicates the expected resonance for an angle between the atom and the laser of 6$^\circ$.  (b) Arrival time distribution obtained from (a) after background count subtraction. (c) Atomic velocity distribution derived from (b) assuming that the number of scattered photons is inversely proportional to the atomic velocity $v$. The velocity distribution peaks at velocities of about $1600$~m/s, i.e. at velocities much lower than the ones found for $^{27}$Al atoms. 
}
\end{center}
\end{figure}

The atomic velocity distribution of laser-ablated calcium atoms can be determined in a similar fashion by analyzing the time-resolved fluorescence of neutral $^{40}$Ca atoms excited on the $^1S_0\leftrightarrow {{^1}P}_1$ transition. As this transition is closed (in contrast to the transition used in the case of $^{27}$Al),  one needs to take into account that the number of photons scattered by an atom is inversely proportional to its velocity. Figure~\ref{CaSpectroscopy} shows the results of such a measurement where the photon arrival time histogram was converted into a velocity histogram that was scaled with velocity to account for the velocity-dependent detection probability of neutral Ca atoms. It can be seen that the atomic velocity distribution peaks at much lower velocities of about $v\approx 1600$~m/s. Similar results were found in previous experiments \cite{Sheridan:2011}.

\section{Discussion and conclusion} 
The simple cooling models developed in this paper predict a decrease of the sympathetic cooling efficiency with the initial ion energy $E_0$, giving rise to cooling times $\tau\sim {E_0}^3$. This points to the need of keeping the initial energy of ions loaded into the trap as low as possible.  

We attribute the long sympathetic cooling times observed in our experiments to the high initial energy of the ions loaded into the trap. Part of this energy is kinetic energy of the laser-ablated atoms; for example, a $^{27}$Al atom with velocity $v=1000$~m/s has a kinetic energy $E_{kin}=0.14$~eV. By pulsing the photo-ionization laser beams following an ablation pulse, slow atoms can be selected from the atomic velocity distribution for photo-ionization. Another contribution is the potential energy of an ion created off-centre. A calcium ion created at $z=500\,\mu$m in a conservative harmonic trap with $\omega_z/(2\pi)=500$~kHz oscillation frequency aquires a potential energy $E_{pot}=m\omega_z^2z^2/2=0.35$~eV. In rf-traps, the dynamical confinement gives rise to additional contributions modifying these energy estimates \cite{Fischer:1959}. Interestingly, the total secular energy $E$ of the ion could be lower than the kinetic energy of the neutral atom being photo-ionized, if the ion could be created at a particular phase of the rf-drive field (by intensity-modulating the photo-ionization laser beam with a sufficiently large bandwidth) and a precisely controlled position that depends on the ion's initial velocity. Within the secular approximation, one finds $\min(E)=E_{kin}/3$ for a pure rf-field, which could result in a substantial reduction of the sympathetic cooling time. 

In our experiments with aluminium, the velocity class selection by delaying the photo-ionization pulse is limited to $v\gtrsim 800$~m/s by the rapidly vanishing tail of slow Al atoms. A second complication is the limited control we have over the positions where atoms become ionized in the trap. As the photo-ionization laser is shown along the axis of the trap, the photo-ionization volume is limited radially by the waist of the laser beam. In the axial direction, ions could be created over a fairly long range limited only by the residual ($z$-position-dependent) Doppler shift of the photo-ionization laser seen by atoms crossing the trap axis off-centre. In experiments with ions created by two-step photo-ionization with two different laser frequencies, the ion creation volume could be reduced by crossed photo-ionization beams. For the slowest available Al atoms, the simple cooling model predicts cooling times on the order of a few seconds. As we never observed cooling times much shorter than one minute, we believe that the initial energy of the ion is dominated by the potential energy it acquires when it is created off-centre. It is currently not clear whether the sympathetic cooling times could be shortened in our experimental setup by restricting the ion creation volume to a region close to the trap centre as in another experiment with  Al$^+$ ions created by laser ablation, shorter cooling times were observed \cite{Chou:2015}. 

In experiments requiring a precise number of ions of different species, long sympathetic cooling times $\tau$ can be a serious obstacle to loading the trap by ablation laser pulses as the success probability $p$ of loading an ion by a single pulse should be much smaller than unity in order to prevent loading more ions than desired. The immediate detection of laser ablation-loaded ions by monitoring the motional state of a cold ion in the trap overcomes this obstacle because it makes it possible to immediately refire the ablation laser if the previous loading trial was unsuccessful. This reduces the average time for loading an ion from $\sim \tau/p$ to $\sim\tau$. Our experiments demonstrate that the motion-dependent coherent electronic excitation of an ion is a very useful diagnostic tool in mixed-crystal experiments, which is applicable not only to cold ions but also to ions with high kinetic energy.

\ack We acknowledge funding by the European Space Agency (contract number 4000102396/10/NL/SFe) and by the European Commission via the integrated project SIQS, and by the Institut f\"ur Quanteninformation GmbH. We thank \'Ad\'am~Selyem for help with the experiment. C.R. would like to thank Jook Walraven and Cornelius Hempel for discussions and helpful comments on the manuscript. 

\appendix
\section{Refined cooling model\label{sec:appendixcoolingmodel}}
The rate of energy loss of the hot ion by the two collisions per oscillation period is given by
\begin{eqnarray}
\frac{dE}{dt}&=&-\frac{\omega}{2\pi}\frac{\langle(\Delta p)^2\rangle}{m_c}\nonumber\\
&=&-\left(\frac{e^2}{4\pi\epsilon_0}\right)^2\frac{m_h^2\omega}{m_c\pi}\left\langle\frac{1}{L^2}\right\rangle
=-\left(\frac{e^2}{4\pi\epsilon_0}\right)^2\frac{m_h^2\omega}{m_c\pi}\int dx\,g_{L}(x)\frac{1}{x^2}.
\label{eq:energylossrateA}
\end{eqnarray}
Here $\langle .\rangle$ indicates an average over the collision parameters which vary from collision to collision as 
the angular momentum is not a conserved quantity. We expect the angular momentum to have a distribution $g_{L}\propto L$ in the interval from $0$ to $\max(L)$ for the following reason: in quantum theory, the operators  $\hat{H}$, $\hat{L}^2$, and $\hat{L}_z$ commute in a three-dimensional central-force potential. For this reason, there is a basis of states with eigenvalues $(E,\hbar^2 l(l+1),\hbar m)$ with $l=0,1,\max(l)=E/(\hbar\omega)-3/2$ and $m=-l,\ldots,l$. Therefore, the state density is linearly increasing with $g_{L}(x)=2(\omega/E)^2x$ up to $L_{max}\approx E/\omega$.

As we consider only small-angle collisions, we also have to introduce a lower cut-off set by the minimum impact parameter $b_{min}=\frac{e^2}{4\pi\epsilon_0 E}$ at which the Coulomb energy equals the energy of the hot ion. This introduces a lower bound of $L_{min}=mvr_{min}$. The integral in~(\ref{eq:energylossrate}) can therefore be replaced by $\int_{L_{min}}^{L_{max}}dxg_L(x)x^{-2}=2L_{max}^{-2}\log(L_{max}/L_{min})$.

In order to write the differential equation~(\ref{eq:energylossrate}) in a compact form, it is convenient to introduce the length and an energy scales already introduced in the main text,
\[
  d=\left(\frac{e^2}{2\pi\epsilon_0m_h\omega^2}\right)^{1/3} \mbox{\hspace{.5cm} and \hspace{.5cm}} E_d = \frac{1}{2}m_h\omega^2d^2,
\]
which correspond to the equilbrium distance of a two-ion crystal and the potential energy of the hot ion at distance $d$. Assuming that $v=\sqrt{2E/m}$, we obtain
\[
\frac{L_{max}}{L_{min}}=\left(\frac{E}{\sqrt[3]{2}E_d}\right)^{3/2}
\]
and the differential equation becomes
\begin{equation}
\frac{dE}{dt}=-\frac{\omega}{2\pi}\frac{8m_h}{m_c}\frac{E_d^3}{E^2}\log\left(\frac{L_{max}}{L_{min}}\right)
=-\frac{\omega}{2\pi}\frac{12m_h}{m_c}\frac{E_d^3}{E^2}\log\left(\frac{E}{\sqrt[3]{2}E_d}\right).
\label{eq:refinedcoolingmodel}
\end{equation}
This differential equation predicts somewhat shorter cooling times than the simple model of~(\ref{eq:energylossrate2}), as shown by the numerical simulations plotted in Figure~\ref{fig:SimulationDifferentMassCase}, but hardly changes the scaling with the initial energy.

The cooling models for both ions of the same and of unequal mass are certainly oversimplistic. They are based on the assumption that the total energy of the initial state is the only parameter relevant for predicting the cooling time. This assumption would be well-justified if the hot ion ergodically explored the phase space during the cooling process. However, the kinetic and potential energy of the hot ion along the direction $i$, $E_i = \frac{1}{2}m_h(\dot{r}_{i}^2+\omega_{h,i}^2r_i^2)$ is nearly a conserved quantity if $E_i\gg E_d$. The time it takes the Coulomb interaction between the ions to redistribute energy  among the different directions is comparable to the time of energy redistribution between the ions. For this reason, the ergodicity assumption is not very good. It would be preferable to have a model that predicts the cooling time as a function of $E_x,E_y,E_z$.

\section{Mapping motional information on the qubit state of an ion\label{sec:appendixcouplingstrength}}
Outside the Lamb-Dicke regime, the coupling strength on the carrier transition depends on the motional state. For an ion in Fock state $|n\rangle$, the Rabi frequency on the carrier is
\begin{equation}
\Omega_{nn}=\Omega_0\langle n|e^{i\eta(a+a^\dagger)}|n\rangle = \Omega_0 L_n(\eta^2)
\end{equation}
where $\Omega_0$ is the Rabi frequency for a ground state cooled ion and $L_n(x)$ a Laguerre polynomial. For an ion in a thermal state of motion the mean squared Rabi frequency
\begin{equation}
\langle \Omega^2\rangle_{\bar{n}} = (1-z)\sum_{n=0}^\infty z^n\Omega_{nn}^2 \mbox{\;\;with\;\;} z=\frac{\bar{n}}{\bar{n}+1}\label{eq:thermalcouplingsquared}
\end{equation}
can be evaluated by using the following relation for the generating function \cite{Erdelyi:1953}
\begin{equation}
\sum_{n=0}^\infty L_n(x)L_n(y)z^n=\frac{1}{1-z}\exp\left(-z\frac{x+y}{1-z}\right)I_0\left(2\frac{\sqrt{xyz}}{1-z}\right)\label{eq:Erdelyi}
\end{equation}
where $I_0$ denotes the modified Bessel function of the first kind. Plugging~(\ref{eq:Erdelyi}) into (\ref{eq:thermalcouplingsquared})
 with $x=y=\eta^2$ yields 
\begin{eqnarray}
\langle\Omega^2\rangle_{\bar{n}}
&=&\Omega_0^2\exp\left(-2\eta^2\bar{n}\right)I_0\left(2\eta^2\bar{n}\sqrt{1+\frac{1}{\bar{n}}}\right)\nonumber\\
&\approx& \Omega_0^2\exp\left(-2\eta^2\bar{n}\right)I_0\left(2\eta^2\bar{n}\right).
\end{eqnarray}
When applying a $\pi$-pulse to an ion in a thermal state, i.e. choosing a pulse duration such that $\Omega_0\tau=\pi$, the probability $p_{\bar{n}}$ of exciting it is given by
\begin{equation}
p_{\bar{n}}= \langle \sin(\Omega_{nn}\tau/2)^2\rangle \approx \sin^2\left(\frac{\pi}{2}\sqrt{\langle \Omega^2\rangle_{\bar{n}}}\right)\label{eq:pofn}
\end{equation}
The last expression is an approximation as higher even moments $\langle \Omega^{2k}\rangle_{\bar{n}}$ with $k=2,3,\ldots$ cannot be written as products of $\langle \Omega^2\rangle_{\bar{n}}$. However, numerical simulations show that $p_{\bar{n}}$ is closely approximated by~(\ref{eq:pofn}).

\section*{References}


\end{document}